\documentclass[12pt]{article}
\usepackage{bm,amsmath,amssymb,graphicx}

\begin{document}
\pagenumbering{arabic}
\begin{titlepage}

\title{A new geometrical approach to metric-affine gravity}

\author{F. F. Faria$\,^{*}$ \\
Centro de Ci\^encias da Natureza, \\
Universidade Estadual do Piau\'i, \\ 
64.002-150 Teresina, PI, Brazil}

\date{}
\maketitle

\begin{abstract}
Here we consider a metric-affine theory of gravity in which the gravitational 
Lagrangian is the scalar curvature. The matter action is allowed to depend also 
on the torsion and the nonmetricity, which are considered as the field 
variables together with the metric. We find the field equations of the theory 
and show that they reduce to Einstein equations in vacuum.  
\end{abstract}

\thispagestyle{empty}
\vfill
\noindent %PACS numbers: 04.20.-q, 04.20.Cv, 04.50.Kd \par
\bigskip
\noindent * fff@uespi.br \par
\end{titlepage}
\newpage

%%%%%%%%%%%%%%%%%%%%%%%%%%%%%%%%%%%%%%%%%%%%%%%%%%%%%%%%%%%%%%%%%%%%%%%%%%%%%%%

\section{Introduction}

It is well known that general relativity (GR) is consistent with several solar 
system experimental tests. However, the standard GR seems unable to explain 
some gravity phenomena on both atomic and cosmological scales. A number of 
alternative theories of gravity have been proposed to explain such phenomena 
but none have been completely successful so far. One of these theories is the 
metric-affine gravity (MAG), which is a generalization of GR based on a 
spacetime with an asymmetric and not metric compatible connection. 

In MAG, the antisymmetric part of the connection is related to torsion 
\cite{Cartan}, whereas the covariant derivative of the metric construct out 
of the connection is related to nonmetricity \cite{Weyl}. The sources of 
torsion and nonmetricity are the matter spin and strain (dilation plus shear) 
currents \cite{Hehl1}, respectively. The geometrical concepts of torsion and 
nonmetricity have concrete physical interpretations as lattice defects in the 
three-dimensional theory of elastic continua \cite{Kroner}.

The field equations of MAG can be derived by means of a metric-affine 
variational principle in which the metric and the connection are considered 
as independent variables. Comparison with the metric variational principle of 
GR suggests the choice of the scalar curvature as the MAG gravitational field 
Lagrangian. The adoption of this Lagrangian together with a matter Lagrangian 
independent of the connection leads to the Einstein field equations of GR 
\cite{Wald}. However, assuming dependence of the matter Lagrangian on the 
connection, which is the most natural choice, yields an unphysical 
MAG model \cite{Hehl2}. The usual procedures to solve such a problem are 
either to generalize the gravitational part of the Lagrangian by adding some 
extra terms \cite{Hehl2}, or to interpret MAG as a gauge theory by introducing 
the coframe ($e^{\mu} = e_{a}\,\!^{\mu}dx^{a}$), besides the metric and the 
connection, as a field variable \cite{Hehl3}.

In this article, we address a new geometrical approach to MAG in which 
we consider the scalar curvature as the gravitational Lagrangian and the 
metric, the torsion and the nonmetricity as the field variables. In 
section 2 we describe the MAG spacetime and the formalism of MAG which we 
shall use throughout this article. In section 3 we derive the MAG field 
equations for the independent metric, torsion and nonmetricity. In section 4 
we analyze the vacuum solution of the MAG field equations. Finally, in section 
5 we present our conclusions.

%%%%%%%%%%%%%%%%%%%%%%%%%%%%%%%%%%%%%%%%%%%%%%%%%%%%%%%%%%%%%%%%%%%%%%%%%%%%%%%

%%%%%%%%%%%%%%%%%%%%%%%%%%%%%%%%%%%%%%%%%%%%%%%%%%%%%%%%%%%%%%%%%%%%%%%%%%%%%%%

\section{MAG spacetime}

%%%%%%%%%%%%%%%%%%%%%%%%%%%%%%%%%%%%%%%%%%%%%%%%%%%%%%%%%%%%%%%%%%%%%%%%%%%%%%%

In GR, gravity is represented by the Riemann spacetime, in which a free 
particle follows the geodesic trajectory
\begin{equation}
\frac{d^2 x^{\lambda}}{d\tau^2}+ 
\stackrel{\circ}{\Gamma}\,\!\!^{\lambda}\,\!\!_{\mu\nu}
\frac{d x^{\mu}}{d\tau}\frac{d x^{\nu}}{d\tau} = 0,
\label{1}
\end{equation}
where
\begin{equation}
\stackrel{\circ}{\Gamma}\,\!\!^{\lambda}\,\!\!_{\mu\nu} = \frac{1}{2} 
g^{\lambda\rho}\left( \partial_{\mu} g_{\rho\nu} + 
\partial_{\nu} g_{\rho\mu}- \partial_{\rho} 
g_{\mu\nu} \right)
\label{2}
\end{equation}
is the Levi-Civita connection, $g_{\mu\nu}$ is the spacetime metric and $\tau$ 
is a suitable affine parameter along the geodesic, which can be taken to 
represent the particle proper time. Both $g_{\mu\nu}$ and 
$\stackrel{\circ}{\Gamma}\,\!\!^{\lambda}\,\!\!_{\mu\nu}$ are symmetric in the 
$\mu\nu$ indices.

The covariant derivative of the metric, the torsion and the curvature construct 
out of the Levi-Civita connection are given by
\begin{equation}
\stackrel{\circ}\nabla\,\!\!\!\!_{\lambda}g_{\mu \nu} = \partial_{\lambda}
g_{\mu \nu} - \stackrel{\circ}{\Gamma}\,\!\!^{\rho}\,\!\!_{\mu\lambda}
g_{\rho\nu} - \stackrel{\circ}{\Gamma}\,\!\!^{\rho}\,\!\!_{\nu\lambda}
g_{\rho\mu} = 0,
\label{3}
\end{equation}
\begin{equation}
\stackrel{\circ}{T}\,\!\!^{\lambda}\,\!\!_{\mu\nu} = 
\stackrel{\circ}{\Gamma}\,\!\!^{\lambda}\,\!\!_{\mu\nu} - 
\stackrel{\circ}{\Gamma}\,\!\!^{\lambda}\,\!\!_{\nu\mu} = 0,
\label{4}
\end{equation}
\begin{equation}
\stackrel{\circ}R\,\!\!^{\alpha}\,\!\!_{\mu\beta\nu}  = \partial_{\beta}\!\!
\stackrel{\circ}{\Gamma}\,\!\!^{\alpha}\,\!\!_{\mu\nu} - \partial_{\nu}\!\! 
\stackrel{\circ}{\Gamma}\,\!\!^{\alpha}\,\!\!_{\mu\beta} + 
\stackrel{\circ}{\Gamma}\,\!\!^{\alpha}\,\!\!_{\sigma \beta}\!\!
\stackrel{\circ}{\Gamma}\,\!\!^{\sigma}\,\!\!_{\mu\nu} - 
\stackrel{\circ}{\Gamma}\,\!\!^{\alpha}\,\!\!_{\sigma \nu}\!
\stackrel{\circ}{\Gamma}\,\!\!^{\sigma}\,\!\!_{\mu\beta},
\label{5}
\end{equation}
respectively. This means that in Riemann spacetime lengths and angles are 
preserved under parallel displacement, infinitesimal parallelograms are 
closed and geodesics deviate from each other. This spacetime geometry 
seem to be fully confirmed experimentally. However, indications from the 
inflationary model \cite{Inflation} and the string theory \cite{String} 
points to a non-Riemannian geometry in the early universe and on the Planck 
scale, respectively. The simplest spacetime having such geometry is MAG 
spacetime. 

The MAG spacetime is endowed with an asymmetric connection 
$\Gamma^{\lambda}\,\!\!_{\mu\nu}$ presenting nonmetricity
\begin{equation}
Q_{\lambda\mu\nu} = - \nabla_{\lambda}g_{\mu \nu} = 
- \partial_{\lambda}g_{\mu \nu} + \Gamma^{\rho}\,\!\!_{\mu \lambda}
g_{\rho\nu} + \Gamma^{\rho}\,\!\!_{\nu\lambda}g_{\rho\mu},
\label{6}
\end{equation}
torsion 
\begin{equation}
T^{\lambda}\,\!\!_{\mu\nu} = \Gamma^{\lambda}\,\!\!_{\mu\nu} - 
\Gamma^{\lambda}\,\!\!_{\nu\mu},
\label{7}
\end{equation}
and curvature
\begin{equation}
R^{\alpha}\,\!\!_{\mu\beta\nu} = \partial_{\beta}
\Gamma^{\alpha}\,\!\!_{\mu\nu}- \partial_{\nu} 
\Gamma^{\alpha}\,\!\!_{\mu\beta}+ \Gamma^{\alpha}\,\!\!_{\sigma 
\beta}\Gamma^{\sigma}\,\!\!_{\mu\nu} - \Gamma^{\alpha}\,\!\!_
{\sigma \nu}\Gamma^{\sigma}\,\!\!_{\mu\beta}.
\label{8}
\end{equation} 
The nonmetricity, which is symmetric in its last two indices, is responsible 
for the change in lengths and angles under parallel displacement, whereas the 
torsion, which is antisymmetric in its last two indices, is responsible for 
the breaking of infinitesimal parallelograms.  The only symmetry of the 
curvature tensor is $R^{\alpha}\,\!\!_{\mu\beta\nu}  = 
- R^{\alpha}\,\!\!_{\mu\nu\beta}$.

In order to get a general expression for the metric-affine connection
$\Gamma^{\lambda}\,\!\!_{\mu \nu}$, we add to equation (6) the same equation with 
the indices $\mu$ and $\lambda$ changed, subtract the same equation with the 
indices $\nu$ and $\lambda$ changed, and find 
\begin{eqnarray} 
Q_{\lambda\mu\nu} + Q_{\mu\lambda\nu} - Q_{\nu\mu\lambda} &=& 
-\partial_{\lambda}g_{\mu \nu} - \partial_{\mu}g_{\lambda \nu}+\partial_{\nu}
g_{\mu \lambda} \nonumber \\ && + \Gamma^{\rho}\,\!\!_{\mu\lambda}g_{\rho\nu}
+\Gamma^{\rho}\,\!\!_{\nu\lambda}g_{\rho\mu}+ \Gamma^{\rho}\,\!\!_{\lambda 
\mu}g_{\rho\nu} \nonumber \\ && +\Gamma^{\rho}\,\!\!_{\nu\mu}g_{\rho\lambda}
-\Gamma^{\rho}\,\!\!_{\mu\nu}g_{\rho \lambda}-\Gamma^{\rho}\,\!\!_{\lambda\nu}
g_{\rho\mu}.
\label{9}
\end{eqnarray}
With some manipulation, we can write this equation as
\begin{equation}
\Gamma^{\lambda}\,\!\!_{\mu\nu} = \, \stackrel{\circ}
{\Gamma}\,\!\!^{\lambda}\,\!\!_{\mu\nu} + 
K^{\lambda}\,\!\!_{\mu\nu} + N^{\lambda}\,\!\!_{\mu\nu} =
\, \stackrel{\circ}{\Gamma}\,\!\!^{\lambda}\,\!\!_{\mu\nu} + 
W^{\lambda}\,\!\!_{\mu\nu},
\label{10}
\end{equation}
where $\stackrel{\circ}{\Gamma}\,\!\!^{\lambda}\,\!\!_{\mu\nu}$ is 
the Levi-Civita connection,
\begin{equation}
K^{\lambda}\,\!\!_{\mu\nu}  = \frac{1}{2} \left(
T^{\lambda}\,\!\!_{\mu\nu} - T_{\mu}\,\!^{\lambda}\,\!\!_{\nu} - 
T_{\nu}\,\!^{\lambda}\,\!\!_{\mu} 
\right)
\label{11}
\end{equation}
is the contortion tensor,
\begin{equation}
N^{\lambda}\,\!\!_{\mu\nu}  = \frac{1}{2} \left( 
Q_{\mu}\,\!^{\lambda}\,\!\!_{\nu}
+Q_{\nu}\,\!^{\lambda}\,\!\!_{\mu}
- Q^{\lambda}\,\!\!_{\mu\nu} \right)
\label{12}
\end{equation}
is the nonmetric part of the connection, and 
\begin{equation}
W^{\lambda}\,\!\!_{\mu\nu} = K^{\lambda}\,\!\!_{\mu\nu} 
+ N^{\lambda}\,\!\!_{\mu\nu}
\label{13}
\end{equation} 
is the distortion tensor. The contortion tensor is antisymmetric in its first 
two indices, the nonmetric part of the connection is symmetric in its last 
two indices and the distortion tensor is asymmetric.

By substituting equation (10) into equation (8), we can write the curvature of 
the metric-affine connection as
\begin{equation}
R^{\alpha}\,\!\!_{\mu\beta\nu} =  
\stackrel{\circ}R\,\!\!^{\alpha}\,\!\!_{\mu\beta\nu} 
+  \stackrel{\circ}\nabla\,\!\!\!\!_{\beta}W^{\alpha}\,\!\!_{\mu \nu} 
- \stackrel{\circ}\nabla\,\!\!\!\!_{\nu}W^{\alpha}\,\!\!_{\mu \beta} 
+ W^{\alpha}\,\!\!_{\rho \beta}W^{\rho}\,\!\!_{\mu \nu} 
- W^{\alpha}\,\!\!_{\rho \nu}W^{\rho}\,\!\!_{\mu\beta},
\label{14}
\end{equation}
where $\stackrel{\circ}R\,\!\!^{\alpha}\,\!\!_{\mu\beta\nu}$ is the 
curvature of GR (Riemann tensor) and $\stackrel{\circ}\nabla\,\!\!\!\!_{\mu}$ 
is the covariant derivative constructed out of the Levi-Civita connection. 
Contracting the curvature tensor (14) in the $\alpha\beta$ indices gives 
the Ricci tensor
\begin{equation}
R_{\mu\nu} = \stackrel{\circ}R_{\mu\nu} 
+  \stackrel{\circ}\nabla\,\!\!\!\!_{\alpha}W^{\alpha}\,\!\!_{\mu \nu} 
- \stackrel{\circ}\nabla\,\!\!\!\!_{\nu}W^{\alpha}\,\!\!_{\mu \alpha} 
+ W^{\alpha}\,\!\!_{\rho \alpha}W^{\rho}\,\!\!_{\mu \nu} 
- W^{\alpha}\,\!\!_{\rho \nu}W^{\rho}\,\!\!_{\mu\alpha},
\label{15}
\end{equation}
which is asymmetric. The scalar curvature $R = g^{\mu\nu}R_{\mu\nu}$ reads
\begin{equation}
R = g^{\mu\nu} \left(\stackrel{\circ}R_{\mu\nu} 
+  \stackrel{\circ}\nabla\,\!\!\!\!_{\alpha}W^{\alpha}\,\!\!_{\mu \nu} 
- \stackrel{\circ}\nabla\,\!\!\!\!_{\nu}W^{\alpha}\,\!\!_{\mu \alpha} 
+ W^{\alpha}\,\!\!_{\rho \alpha}W^{\rho}\,\!\!_{\mu \nu} 
- W^{\alpha}\,\!\!_{\rho \nu}W^{\rho}\,\!\!_{\mu\alpha}\right).
\label{16}
\end{equation}
Note that is possible to construct another Ricci tensor $R'_{\mu\nu}$ by 
contracting the curvature tensor (14) in its first two indices. However, the 
scalar curvature is uniquely defined by equation (16), since $g_{\mu\nu}$ is 
symmetric and $R'_{\mu\nu}$ is antisymmetric.

%%%%%%%%%%%%%%%%%%%%%%%%%%%%%%%%%%%%%%%%%%%%%%%%%%%%%%%%%%%%%%%%%%%%%%%%%%%%%%%

\section{MAG field equations}

%%%%%%%%%%%%%%%%%%%%%%%%%%%%%%%%%%%%%%%%%%%%%%%%%%%%%%%%%%%%%%%%%%%%%%%%%%%%%%%

The MAG field equations can be obtained by the variational principle  
\begin{equation}
\delta S_{g} + \delta S_{m} = 0,
\label{17}
\end{equation}
where $S_{g}$ is the action of the gravitational field and $S_{m}$ is the 
action of the matter field, which includes matter and all fields that interact
with the gravitational field.

The MAG gravitational action adopted here is given by
\begin{equation}
S_{g} = - \frac{1}{2kc}\int{d^{4}x\sqrt{-g} R},
\label{18}
\end{equation}
where $k = 8\pi G/c^4$ ($G$ is the gravitational constant and $c$ is the 
speed of light in vacuum) and $R = R\left(g, T, Q\right)$. Using equation (16) 
and neglecting surface terms (which do not contribute to the field equations), 
we find that equation (18) reduces to  
\begin{equation}
S_{g} = - \frac{1}{2kc} \int  d^{4}x\sqrt{-g} g^{\mu\nu}\biggl[  
\stackrel{\circ}R_{\mu\nu} + W^{\alpha}\,\!\!_{\rho \alpha}
W^{\rho}\,\!\!_{\mu \nu} - 
W^{\alpha}\,\!\!_{\rho \nu}W^{\rho}\,\!\!_{\mu\alpha} \biggl].
\label{19}
\end{equation}

Varying the gravitational action (19) with respect to the metric tensor, 
contortion tensor (which is equivalent to varying with respect to the torsion 
tensor) and the nonmetric part of the connection (which is equivalent to 
varying with respect to the nonmetricity tensor) gives
\begin{eqnarray}
 \delta S_{g} &=& - \frac{1}{2kc} \int  d^{4}x\sqrt{-g} \Biggl[ \biggl( 
\stackrel{\circ}R_{\mu\nu}  - \frac{1}{2}g_{\mu\nu}\!\!\stackrel{\circ}R 
- kU_{\mu\nu}\biggl) \delta g^{\mu\nu} \nonumber \\ && +
\biggl( \delta^{\lambda}\,\!\!_{[\mu}{}W_{\nu]\rho}{}\,\!^{\rho} 
+ \delta^{\lambda}_{\phantom{\lambda}[\nu}W^{\rho}_{\phantom{\rho}\mu]\rho} 
- W^{\lambda}\,\!\!_{[\mu\nu]} - W_{[\nu}{}\,\!\!^{\lambda}\,\!\!_{\mu]}
\biggl) \delta K^{\mu\nu}\,\!\!_{\lambda} 
\nonumber \\ && + \biggl( \delta^{\lambda}\,\!\!_{(\mu}{}
W_{\nu)\rho}{}\,\!^{\rho} + g_{\mu\nu}W^{\rho\lambda}\,\!\!_{\rho} 
- W_{(\mu\nu)}\,\!\!^{\lambda} - W_{(\mu}{}\,\!^{\lambda}\,\!\!_{\nu)}{} 
\biggl) \delta N_{\lambda}\,\!^{\mu\nu} \Biggr],
\label{20}
\end{eqnarray}
where 
\begin{equation}
U_{\mu\nu} = \frac{1}{k} \Biggl[W^{\alpha}\,\!\!_{\rho (\nu}{}
W^{\rho}\,\!\!_{\mu)\alpha} - W^{\alpha}\,\!\!_{\rho\alpha}
W^{\rho}\,\!\!_{(\mu\nu)} - \frac{1}{2}g_{\mu\nu}\biggl( 
W^{\alpha}\,\!\!_{\rho \sigma}W^{\rho\sigma}\,\!\!_{\alpha} 
- W^{\alpha}\,\!\!_{\rho\alpha}W^{\rho\sigma}\,\!\!_{\sigma} 
\biggr) \Biggr].
\label{21}
\end{equation}
The symbols $( \ )$ and $[ \ ]$ around the indices denote symmetrization and 
antisymmetrization, respectively. For example, $A_{(\mu\nu)} = 
\frac{1}{2}(A_{\mu\nu}+A_{\nu\mu})$ and $A_{[\mu\nu]} = \frac{1}{2}
(A_{\mu\nu}-A_{\nu\mu})$. 

The MAG matter action that we consider is given by
\begin{equation}
S_{m} = \frac{1}{c}\int{d^{4}x\mathcal{L}_{m}}, 
\label{22}
\end{equation}
where $\mathcal{L}_{m} = \mathcal{L}_{m}\left(\Psi, g, T, Q\right)$ is the 
Lagrangian density of the matter field $\Psi$. The variation of the matter 
action (22) with respect to the metric tensor, contortion tensor and the 
nonmetric part of the connection reads
\begin{equation}
\delta S_{m} = \frac{1}{2c}\int{d^{4}x\sqrt{-g} \Biggl[ T_{\mu\nu} \, 
\delta g^{\mu\nu} +  S_{\mu\nu}\,\!\!^{\lambda} 
\, \delta K^{\mu\nu}\,\!\!_{\lambda} +  \Sigma^{\lambda}\,\!\!_{\mu\nu} 
\, \delta N_{\lambda}\,\!^{\mu\nu}\Biggr]},
\label{23}
\end{equation}
where 
\begin{equation}
T_{\mu\nu} = \frac{2}{\sqrt{-g}} \frac{\delta \mathcal{L}_{m}}{\delta 
g^{\mu\nu}}
\label{24}
\end{equation}
is the matter energy-momentum tensor,
\begin{equation}
S_{\mu\nu}\,\!\!^{\lambda} = \frac{2}{\sqrt{-g}} \frac{
\delta \mathcal{L}_{m}}{\delta K^{\mu\nu}\,\!\!_{\lambda}}
\label{25}
\end{equation}
is the matter spin tensor, and
\begin{equation}
\Sigma^{\lambda}\,\!\!_{\mu\nu} = \frac{2}{\sqrt{-g}} \frac{
\delta \mathcal{L}_{m}}{\delta N_{\lambda}\,\!^{\mu\nu}}
\label{26}
\end{equation}
is the matter strain tensor, which can be split into a trace part (dilation) 
\begin{equation}
\not\!{\Sigma}^{\lambda}\,\!\!_{\mu\nu} = \frac{1}{4}g_{\mu\nu}
\Sigma^{\lambda\rho}\,\!\!_{\rho},
\label{27}
\end{equation}
and a traceless part (shear)
\begin{equation}
\widehat{\Sigma}^{\lambda}\,\!\!_{\mu\nu} = \Sigma^{\lambda}\,\!\!_{\mu\nu} 
- \frac{1}{4}g_{\mu\nu}\Sigma^{\lambda\rho}\,\!\!_{\rho}.
\label{28}
\end{equation}

Note that the matter energy-momentum tensor is symetric, the matter spin tensor 
is antisymmetric in its first two indices and the matter strain tensor is 
symmetric in its last two indices. An example of a physical matter source of 
MAG is the hyperfluid \cite{Obukhov}, which is a classical model of a 
continuous medium with energy-momentum, spin and strain. A different model of 
a perfect fluid with the same properties was proposed in \cite{Smalley}.

From equations (17), (20) and (23), we find the MAG field equations
\begin{equation}
\stackrel{\circ}R_{\mu\nu}  - \frac{1}{2}g_{\mu\nu}\stackrel{\circ}R \,
= k \left(T_{\mu\nu} + U_{\mu\nu} \right),
\label{29}
\end{equation}
\begin{equation}
\delta^{\lambda}\,\!\!_{[\mu}{}W_{\nu]\rho}{}\,\!^{\rho} + 
\delta^{\lambda}_{\phantom{\lambda}[\nu}W^{\rho}_{\phantom{\rho}\mu]\rho}
- W^{\lambda}\,\!\!_{[\mu\nu]} - W_{[\nu}{}\,\!\!^{\lambda}\,\!\!_{\mu]}
 = kS_{\mu\nu}\,\!\!^{\lambda},
\label{30}
\end{equation}
\begin{equation}
\delta^{\lambda}\,\!\!_{(\mu}{}W_{\nu)\rho}{}\,\!^{\rho} 
+ g_{\mu\nu}W^{\rho\lambda}\,\!\!_{\rho}  
- W_{(\mu\nu)}\,\!\!^{\lambda} - W_{(\mu}{}\,\!^{\lambda}\,\!\!_{\nu)}{}
= k\Sigma^{\lambda}\,\!\!_{\mu\nu}.
\label{31}
\end{equation}
The first MAG field equations (29) are a generalization of the Einstein 
equations with a correction to the energy-momentum tensor from the spin 
and strain contributions to the spacetime geometry. The second MAG field 
equations (30) and the third MAG field equations (31) are algebraic relations 
linking spin and strain with torsion and nonmetricity. Note that even though 
the torsion and the nonmetricity do not propagate outside the matter, the spin 
and the strain of the matter influence the geometry also outside the matter 
through the metric tensor.  

%%%%%%%%%%%%%%%%%%%%%%%%%%%%%%%%%%%%%%%%%%%%%%%%%%%%%%%%%%%%%%%%%%%%%%%%%%%%%%%

\section{Vacuum solution}

%%%%%%%%%%%%%%%%%%%%%%%%%%%%%%%%%%%%%%%%%%%%%%%%%%%%%%%%%%%%%%%%%%%%%%%%%%%%%%%

In order to obtain the vacuum solution to MAG field equations, we have to 
resolve the combined system of algebraic equations (30) and (31). Taking into 
account equations (11-13), we can rewrite equations (30-31) as
\begin{equation}
T^{\lambda}\,\!\!_{\nu\mu} + 2\delta^{\lambda}\,\!\!_{[\mu}{}T_{\nu]} 
+ Q_{[\nu\mu]}\,\!\!^{\lambda} + \delta^{\lambda}\,\!\!_{[\mu}{}
Q^{\rho}\,\!\!_{\rho\nu]} - \delta^{\lambda}\,\!\!_{[\mu}{}Q_{\nu]} 
= kS_{\mu\nu}\,\!\!^{\lambda},
\label{32}
\end{equation}
\begin{equation}
T_{(\mu\nu)}\,\!\!^{\lambda} + \delta^{\lambda}\,\!\!_{(\mu}{}T_{\nu)} 
- g_{\mu\nu}T^{\lambda} - Q^{\lambda}\,\!\!_{\mu\nu} 
+ \delta^{\lambda}\,\!\!_{(\mu}{}Q^{\rho}\,\!\!_{\rho\nu)} 
- \frac{1}{2}\delta^{\lambda}\,\!\!_{(\mu}{}Q_{\nu)} 
+ \frac{1}{2}g_{\mu\nu}Q^{\lambda} 
= k\Sigma^{\lambda}\,\!\!_{\mu\nu},
\label{33}
\end{equation}
where $T^{\lambda} = T_{\rho}\,\!^{\rho\lambda}$ is the trace of the torsion 
tensor and $Q^{\lambda} = Q^{\lambda\rho}\,\!\!_{\rho}$ is the Weyl vector. 

With some manipulation, we find that equation (32) is equivalent to
\begin{equation}
T^{\lambda}\,\!\!_{\nu\mu} + Q_{[\nu\mu]}\,\!\!^{\lambda} = k\left(
S_{\mu\nu}\,\!\!^{\lambda} + \delta^{\lambda}\,\!\!_{[\mu}{}
S_{\nu]\rho}\,\!\!^{\rho}\right),
\label{34}
\end{equation}
\begin{equation}
T^{\lambda}\,\!\!_{\nu\mu} + 2T_{[\nu\mu]}\,\!\!^{\lambda}  
 = k\left(S_{\mu\nu}\,\!\!^{\lambda} 
+ 2S^{\lambda}\,\!\!_{[\mu\nu]}\right).
\label{35}
\end{equation}
Similarly, we can see that equation (33) is equivalent to
\begin{eqnarray}
&&Q^{\lambda}\,\!\!_{\mu\nu} - T_{(\mu\nu)}\,\!\!^{\lambda} 
- \frac{1}{3}g_{\mu\nu}T^{\lambda}
- \frac{1}{3}\delta^{\lambda}\,\!\!_{(\mu}{}T_{\nu)} 
= k\biggl[-\Sigma^{\lambda}\,\!\!_{\mu\nu} 
+ \frac{1}{2}g_{\mu\nu}\left( \Sigma^{\lambda\rho}\,\!\!_{\rho} 
- \frac{2}{3}\Sigma^{\rho\lambda}\,\!\!_{\rho}\right) \nonumber \\ &&
+ \frac{2}{3}\delta^{\lambda}\,\!\!_{(\mu}{}\Sigma^{\rho}\,\!\!_{\nu)\rho} 
\biggr],
\label{36}
\end{eqnarray}
\begin{eqnarray}
&&Q^{\lambda}\,\!\!_{\mu\nu} +2Q_{(\mu\nu)}\,\!\!^{\lambda} 
- \frac{2}{3}g_{\mu\nu}T^{\lambda}
- \frac{4}{3}\delta^{\lambda}\,\!\!_{(\mu}{}T_{\nu)} 
 = k\biggl[-\Sigma^{\lambda}\,\!\!_{\mu\nu} 
 + \frac{1}{2}g_{\mu\nu}\left( \Sigma^{\lambda\rho}\,\!\!_{\rho} 
- \frac{2}{3}\Sigma^{\rho\lambda}\,\!\!_{\rho}\right) \nonumber \\ &&
+\frac{2}{3}\delta^{\lambda}\,\!\!_{(\mu}{}\Sigma^{\rho}\,\!\!_{\nu)\rho}
+ \delta^{\lambda}\,\!\!_{(\mu}{}\Sigma_{\nu)\rho}\,\!\!^{\rho}
-2\Sigma_{(\mu\nu)}\,\!\!^{\lambda}\biggr].
\label{37}
\end{eqnarray}
In vacuum, where $T_{\mu\nu} = S_{\mu\nu}\,\!\!^{\lambda} 
= \Sigma^{\lambda}\,\!\!_{\mu\nu} = 0$, equations (34-37) reduces to
\begin{equation}
T^{\lambda}\,\!\!_{\nu\mu} + Q_{[\nu\mu]}\,\!\!^{\lambda} = 0,
\label{38}
\end{equation}
\begin{equation}
T^{\lambda}\,\!\!_{\nu\mu} + 2T_{[\nu\mu]}\,\!\!^{\lambda} =0,
\label{39}
\end{equation}
\begin{equation}
Q^{\lambda}\,\!\!_{\mu\nu} - T_{(\mu\nu)}\,\!\!^{\lambda} 
- \frac{1}{3}g_{\mu\nu}T^{\lambda}
- \frac{1}{3}\delta^{\lambda}\,\!\!_{(\mu}{}T_{\nu)} = 0,
\label{40}
\end{equation}
\begin{equation}
Q^{\lambda}\,\!\!_{\mu\nu} +2Q_{(\mu\nu)}\,\!\!^{\lambda} 
- \frac{2}{3}g_{\mu\nu}T^{\lambda}
- \frac{4}{3}\delta^{\lambda}\,\!\!_{(\mu}{}T_{\nu)} =0.
\label{41}
\end{equation}

By contracting the $\lambda\nu$ indices in equations (38-41), we obtain
\begin{equation}
T_{\mu} + \frac{1}{2} Q^{\rho}\,\!\!_{\rho\mu} 
- \frac{1}{2}Q_{\mu} = 0,
\label{42}
\end{equation}
\begin{equation}
Q^{\rho}\,\!\!_{\rho\mu} - \frac{2}{3}T_{\mu} = 0,
\label{43}
\end{equation}
\begin{equation}
2Q^{\rho}\,\!\!_{\rho\mu} + Q_{\mu} - 4T_{\mu}  = 0,
\label{44}
\end{equation}
whose only solution is
\begin{equation}
T_{\mu} = Q_{\mu} = Q^{\rho}\,\!\!_{\rho\mu} = 0.
\label{45}
\end{equation}
Thus equations (40-41) reduces to
\begin{equation}
Q^{\lambda}\,\!\!_{\mu\nu} - T_{(\mu\nu)}\,\!\!^{\lambda} = 0,
\label{46}
\end{equation}
\begin{equation}
Q^{\lambda}\,\!\!_{\mu\nu} + 2Q_{(\mu\nu)}\,\!\!^{\lambda} = 0.
\label{47}
\end{equation}

Combining equations (38-39) and equations (46-47) gives
\begin{equation}
T^{\lambda}\,\!\!_{\mu\nu} = Q^{\lambda}\,\!\!_{\mu\nu} = 0,
\label{48}
\end{equation}
and, consequently, $U_{\mu\nu} = 0$. Therefore, in vacuum, the first MAG 
field equations (29) reduces to the Einstein equations
\begin{equation}
\stackrel{\circ}R_{\mu\nu}  - \frac{1}{2}g_{\mu\nu}\stackrel{\circ}R \,= 0.
\label{49}
\end{equation}
Thus we conclude that  MAG is equivalent to GR in vacuum.

%%%%%%%%%%%%%%%%%%%%%%%%%%%%%%%%%%%%%%%%%%%%%%%%%%%%%%%%%%%%%%%%%%%%%%%%%%%%%%%

\section{Final remarks}

%%%%%%%%%%%%%%%%%%%%%%%%%%%%%%%%%%%%%%%%%%%%%%%%%%%%%%%%%%%%%%%%%%%%%%%%%%%%%%%

In this article, we proved that is possible obtain a physical MAG model by the 
variational principle with the scalar curvature as the gravitational 
Lagrangian. In most MAG approaches, the gravitational action and the matter 
action are varied with respect to the metric and the connection. However, the 
variation of the total action (with the scalar curvature as the gravitational 
Lagrangian) with respect to the connection constrains possible 
forms of matter that the theory can describe leading to inconsistent 
field equations. We solved this problem by varying the gravitational action 
construct out of the scalar curvature and the matter action with respect to 
the torsion and the nonmetricity instead of the connection. As a result we 
have obtained consistent field equations which reduces to Einstein equations 
in vacuum. It should be important to investigate the solutions of these field 
equations in matter with spin and strain. This question is under consideration 
now.

%%%%%%%%%%%%%%%%%%%%%%%%%%%%%%%%%%%%%%%%%%%%%%%%%%%%%%%%%%%%%%%%%%%%%%%%%%%%%%%
\section*{Acknowledgments}

 The author would like to thank J. B. Fromiga for useful discussions.

%%%%%%%%%%%%%%%%%%%%%%%%%%%%%%%%%%%%%%%%%%%%%%%%%%%%%%%%%%%%%%%%%%%%%%%%%%%%%%%

%%%%%%%%%%%%%%%%%%%%%%%%%%%%%%%%%%%%%%%%%%%%%%%%%%%%%%%%%%%%%%%%%%%%%%%%%%%%%%%

\end{document}